\documentclass[prd, twocolumn, superscriptaddress, nofootinbib,
  showpacs,10pt]{revtex4}
\usepackage{graphicx,amsmath,amsfonts,amssymb,dcolumn,epsfig,bm}

\expandafter\ifx\csname package@font\endcsname\relax\else
 \expandafter\expandafter
 \expandafter\usepackage
 \expandafter\expandafter
 \expandafter{\csname package@font\endcsname}%
\fi
\begin{document}

\title{Classicalization of inflationary perturbations by collapse models in the light of BICEP2}%
\author{Suratna Das}%
\email{suratna@tifr.res.in}
\author{Satyabrata Sahu}%
\email{satyabrata@tifr.res.in}%
\author{Shreya Banerjee}
\email{shreya.banerjee@tifr.res.in}
\author{T. P. Singh}%
\email{tpsingh@tifr.res.in}
\affiliation{Tata Institute of Fundamental Research, Mumbai 400005, India}

\begin{abstract}
Inflationary and hence quantum origin of primordial perturbations is
on a firmer ground than ever post the BICEP2 observations of
primordial gravitational waves. One crucial ingredient of success of
this paradigm rests on explaining the observed classicality of
cosmological inhomogeinities despite their quantum origin. Although
decoherence provides a partial understanding of this issue, the
question of single outcome motivates the analysis of quantum collapse
models in cosmological context which generically modify the dynamics
of primordial perturbations and hence can leave their imprints on
observables. We revisit one such recently proposed working model of
classicalization by spontaneous collapse \cite{Das:2013qwa} in the
light of BICEP2 observations to look for possible modifications to
tensor power spectra and their implications. We show that it can
potentially change the consistency relation of single-field models and
a precise measurement of $n_T$ and its running could serve as a test
of such dynamics in the early universe.
\pacs{98.80.Cq, 11.10.Lm, 98.70.Vc}
\end{abstract}
\maketitle
\section{Introduction}

The recent observation of $B-$mode polarization of Cosmic Microwave
Background Radiation (CMBR) by BICEP2 \cite{Ade:2014xna} has put the
cosmological inflationary paradigm \cite{Guth:1980zm, Linde:1983gd} on
stronger footing than ever by confirming one of its many observational
predictions, namely the generation of primordial tensor modes or
primordial gravitational waves. Inflationary dynamics gives rise to
both scalar and tensor fluctuations during inflation which then
`redshift' out of the horizon and freeze. These primordial scalar
fluctuations, up on re-entering the horizon at later stages, give rise
to $TT$ anisotropy spectrum of the CMBR that has been measured by many
observations such as WMAP \cite{Hinshaw:2012aka} and PLANCK
\cite{Ade:2013zuv}. On the other hand, the tensor modes give rise to
the $BB$ spectrum of CMBR which has recently been detected by BICEP2
\cite{Ade:2014xna}.

Detection of $B-$modes of CMBR is of profound importance as according
to the simplest model of inflation it sets the scale of inflation in a
unique way, which turns out to be surprisingly close to the Grand
Unification (GUT) scale ($\sim 10^{16}$ GeV). But this is not all what
the detection of $B-$modes by BICEP2 can provide. BICEP2 measurement
has further reinforced the quantum nature of gravity as it indicates
that the tensor perturbations generated during inflation are of
quantum nature. BICEP2 has measured the power of these tensor modes
over the scalar ones on large cosmological scales, called the
tensor-to-scalar ratio $r$ and the measured value of $r$ by BICEP2 is
\cite{Ade:2014xna}
\begin{eqnarray}
r=0.2^{+0.07}_{-0.05}\,.
\end{eqnarray}
It has been argued in \cite{Ashoorioon:2012kh} that if these
primordial gravitational waves were of classical nature, it would have
produced negligible tensor amplitude compared to the scalar one and
the value of $r$ would have at least been suppressed by an extra power
of slow-roll parameter $\epsilon$ in comparison to the standard
scenario. That these primordial tensor modes are indeed of quantum
nature has also been argued in other literature such as
\cite{Bose:2001fa, Krauss:2013pha, Krauss:2014sua}. \footnote{However,
  it has been pointed out that large enough primordial gravitational
  wave signal can be produced via purely classical mechanisms, such as
  gravitational bremsstrahlung, in a multi-field scenario
  \cite{Senatore:2011sp} or non-perturbatively generated gravitational
  waves via electromagnegtic fields amplified by an axion-like inflaton
  \cite{Cook:2011hg} which can overshadow the quantum tensor signal.}
Primordial scalar perturbations are also generated quantum
mechanically and this scenario is well supported by observations.

From this discussion it is evident that the primordial modes, both
scalar and tensor, are being generated quantum mechanically at very
early times. But these quantum fluctuations are the ones which give
rise to CMBR temperature fluctuations and its $B-$mode polarization on
large angular scales which are classical in nature. This leads us to
the problem of quantum-to-classical transition in the cosmological
context, which is a more serious form of the so-called ``quantum
measurement problem'' which also prevails in laboratory systems.

Heuristically, the classical nature of primordial perturbations is
argued by the large occupation number of superhorizon modes and
effective irrelevance of the commutator of the field variable and its
conjugate momentum on superhorizon scales. However, the perturbations
generically evolve into highly squeezed states \cite{Polarski:1995jg,
  Bose:2001fa, Albrecht:1992kf} on superhorizon scales, which are
highly non-classical states. But the quantum nature is not directly
evident in observations because it can be shown that quantum
expectations of highly squeezed states are indistinguishable from
average of a classical stochastic field. This is so-called
`decoherence without decoherence' \cite{Polarski:1995jg}. Furthermore
decoherence, despite the ambiguity in system and environment spilt for
a cosmological scenario, selects the field amplitude basis as the
pointer basis of the system and justifies the standard calculation. At
this point, what still remains unresolved is the issue of single
outcome. This problem is also present in laboratory systems and only
gets more intriguing in cosmological context
\cite{Martin:2012pea}. One possible way out is to appeal to the
many-worlds interpretation. The other alternative is the so-called
collapse models which we explore in this paper. Just as in laboratory
systems, it is important to investigate whether and how collapse
models can be distinguished from the standard quantum mechanical setup
in a cosmological context.

In a generic collapse model Schr\"{o}dinger equation is modified by
adding stochastic and non-linear terms. The stochastic nature of the
equation helps explaining the probabilistic outcome of quantum
measurements without allowing for superluminal communication and the
presence of non-linear terms breaks down the underlying superposition
principle of quantum mechanics. For a detailed review on collapse
models refer to \cite{Bassi:2012bg}. Though a proper field theoretic
treatment of collapse dynamics is not yet known, a few attempts have
been made to apply such collapse models, especially Continuous
Spontaneous Localization (CSL) model \cite{Bassi:2012bg}, into
inflationary dynamics to resolve the problem of quantum to classical
transition in cosmological context, such as \cite{Martin:2012pea},
\cite{Das:2013qwa}, \cite{Canate:2013isa}. As these collapse models
explicitly modify the standard dynamics, we can anticipate that its
observational implications would diverge from the standard dynamics
and the aim of this brief paper is to determine how and where these
collapse mechanisms differ from the standard scenario observationally
within the context of an illustrative example in the light of recent
BICEP observation.
\section{Generic Single-field slow-roll inflation and its observational implications}

The most economic model of inflation is the so-called slow-roll single
field model. The power spectra for scalar and tensor perturbations in
such a model can be given as \cite{Baumann:2009ds}
\begin{eqnarray}
{\mathcal P}_{\mathcal R}&=&\frac{1}{8\pi^2}\left(\frac{H^2}{\epsilon M_{\rm Pl}^2}\right)\left(\frac{k}{k_*}\right)^{n_s-1}\equiv A_s(k_*)\left(\frac{k}{k_*}\right)^{n_s-1},\nonumber\\
{\mathcal P}_{h}&=&\frac{2}{\pi^2}\left(\frac{H^2}{M_{\rm Pl}^2}\right)\left(\frac{k}{k_*}\right)^{n_T}\equiv A_T(k_*)\left(\frac{k}{k_*}\right)^{n_T}
\label{spectra}
\end{eqnarray}
respectively, where $k_*$ is called the pivot scale and for PLANCK it
is chosen to be $k_*=0.05$ Mpc$^{-1}$. It is to be noted that as the
comoving curvature perturbations, denoted by $\mathcal R$, and the
tensor modes, denoted by $h$, freeze on superhorizon scales, it is
customary to derive the power spectra given above at horizon crossing
of each mode $(k=aH)$ during inflation. In the above equations $H$ is
the Hubble parameter during inflation, $M_{\rm Pl}=2.4\times10^{18}$
GeV is the reduced Planck mass, $\epsilon$ is the first Hubble
slow-roll parameter defined as 
\begin{eqnarray}
\epsilon\equiv-\frac{\dot H}{H^2},
\end{eqnarray}
$n_s$ and $n_T$ are the scalar spectral index and the tensor spectral
index respectively which are the measures of the scale dependence of
the respective spectrum and for this generic slow-roll single-field
model turn out to be
\begin{eqnarray}
n_s-1&=&2\eta-4\epsilon,\nonumber\\
n_T&=&-2\epsilon,
\label{indices}
\end{eqnarray}
where $\eta$ is the second Hubble slow-roll parameter defined as
\begin{eqnarray}
\eta\equiv-\frac{\ddot\varphi_0}{H\dot\varphi_0},
\end{eqnarray}
where $\varphi_0$ is the inflaton field. 

Each of the tensor modes, which are the traceless and transverse part
of the metric fluctuations, is associated with two helicity states,
often denoted as + and $\times$ polarization. As has been first
pointed out by Grishchuk in \cite{Grishchuk:1974ny}, the evolution of
the Fourier modes of each of the helicity states of these tensor modes
is identical to that of a massless scalar in de Sitter space with the
correspondence
\begin{eqnarray}
h_k^s=\frac{2}{aM_{\rm Pl}}v_k^s,
\label{tensor-conf}
\end{eqnarray}
where $s=+,\times$ is the helicity states and $v_k^s$ is the
re-defined tensor modes. Hence the evolution of quantum tensor modes
can be reduced to that of two decoupled massless scalar modes. We will
make use of this fact while dealing with CSL-modified inflationary
dynamics.

Now, let us consider the observational implications of this simplest
model of inflation. PLANCK measures the amplitude $A_s$ and the scalar
spectral index $n_s$ as \cite{Ade:2013zuv}
\begin{eqnarray}
A_s&=&2.215\times10^{-9},\nonumber\\
n_s&=&0.9603\pm0.0073,
\end{eqnarray}
from the $TT$ anisotropy spectrum of the CMBR. BICEP2, by detecting
the $B-$polarization of CMBR measured tensor-to-scalar ratio $r$,
defined as
\begin{eqnarray}
r\equiv\frac{A_T}{A_s},
\end{eqnarray}
to be $0.2$. According to the simplest single field model the
tensor-to-scalar ratio turns out to be of the order of slow-roll
parameter:
\begin{eqnarray}
r=16\epsilon,
\end{eqnarray}
which then yields the consistency relation of this single-field model
given as
\begin{eqnarray}
r=-8n_T.
\end{eqnarray}
Thus independent measurements of $r$ and $n_T$ can unambiguously
determine whether the inflationary dynamics was indeed that simple or
not.

The recent observation of $r$ by BICEP2 and the measurement of scalar
amplitude $A_s$ by PLANCK indirectly provides the amplitude of the
tensor perturbations. But one can see from Eq.~(\ref{spectra}) that
the amplitude of the tensor power solely depends on the Hubble
parameter during inflation and considering the central values of
observed $r$ and $A_s$ one gets the Hubble parameter during inflation
as
\begin{eqnarray}
H\simeq1.1\times10^{14}\,{\rm GeV}.
\end{eqnarray}
Now, during inflationary era the universe is dominated by the
potential energy of the inflaton field and thus the Friedmann equation
during inflation is written as
\begin{eqnarray}
H^2=\frac{1}{3M_{\rm Pl^2}}V(\varphi_0).
\end{eqnarray}
Hence the BICEP2 measurement of $r$ also sets the scale of inflation as 
\begin{eqnarray}
V^{1/4}\sim2.1\times10^{16}\,{\rm GeV},
\end{eqnarray}
which is very close to the GUT scale. However, inferring the scale of
inflation from tensor amplitude becomes more involved if one invokes
large extra dimensions \cite{Ho:2014xza} or classical sources of
primordial gravitational waves in a multi-field scenario
\cite{Senatore:2011sp}.

Another implication of the BICEP2 result comes from the Lyth bound
\cite{Lyth:1996im}. Writing the first slow-roll parameters $\epsilon$
as
\begin{eqnarray}
\epsilon=\frac{M_{\rm Pl}^2}{2}\frac{\dot\varphi_0^2}{H^2},
\end{eqnarray}
one can determine the field excursion during the inflation as
\begin{eqnarray}
\Delta\varphi_0={\mathcal O}(1)\left(\frac{r}{0.01}\right)^{\frac12}M_{\rm Pl},
\end{eqnarray}
which shows that the inflaton field excursion is super-Planckian
during inflation if $r=0.2$, as has been observed by BICEP2. This
leads to some tension with effective field theory description of
inflation as the field excursion becomes of the same order of the
natural cutoff scale \footnote{However there is a recent debate in
  literature whether sub-Planckian field excursion can be made
  consistent with recent BICEP2 observation in a single-field scenario
  \cite{Choudhury:2014kma,Antusch:2014cpa}}.

\section{CSL-modified single-field dynamics and its observational implications}

\subsection{Scalar perturbations}
For this analysis we would keep our focus on the derivations done in
\cite{Martin:2012pea, Das:2013qwa}. The collapse models modify the
dynamics of particles in the Schr\"{o}dinger picture. Hence while
applying the CSL modifications to the inflationary dynamics one
requires to analyze the field dynamics in the Schr\"{o}dinger
picture. An elaborate analysis of Schr\"{o}dinger picture dynamics of
scalar perturbations during inflation is given in
\cite{Polarski:1995jg} (see also \cite{Sriramkumar:2004pj}). We would
study the primordial scalar perturbations during inflation in terms of
the so-called Mukhanov-Sakasi variable, a gauge-invariant quantity
denoted by $\zeta$ , which is related to the comoving curvature
perturbation as
\begin{eqnarray}
\zeta(\tau,\mathbf{x})=\frac{a\varphi_0'}{\mathcal{H}}\mathcal{R}(\tau,\mathbf{x}).
\label{ms-R}
\end{eqnarray}
In Schr\"{o}dinger picture, the standard scalar perturbation is
analyzed in terms of its wave-functional \cite{Polarski:1995jg}
defined as
\begin{eqnarray}
\Psi\left[\zeta(\tau,\mathbf x)\right]=\prod_{\mathbf k}\Psi_{\mathbf
  k}^{\rm R}\left[\zeta^{\rm R}_{\mathbf k}(\tau)\right]\Psi_{\mathbf
  k}^{\rm I}\left[\zeta^{\rm I}_{\mathbf k}(\tau)\right]
\label{scalar-func}
\end{eqnarray}
where we have
\begin{eqnarray}
\zeta_{\mathbf  k}(\tau)=\frac{1}{\sqrt{2}}\left(\zeta_{\mathbf  k}^{\rm R}(\tau)+i\zeta^{\rm I}_{\mathbf  k}(\tau)\right).
\end{eqnarray}
These wave functionals satisfy the functional Schr\"{o}dinger equation
as
\begin{eqnarray}
i\frac{\partial \Psi_{\mathbf k}^{\rm
  R,I}}{\partial\tau}=\hat{\mathcal H}^{\rm R,I}_{\mathbf k}\Psi^{\rm R,I}_{\mathbf k},
\label{sch-eq}
\end{eqnarray}
where the Hamiltonian $\hat{\mathcal H}_{\mathbf k}\equiv
\hat{\mathcal H}^{\rm R}_{\mathbf k}+\hat{\mathcal H}^{\rm I}_{\mathbf
  k}$ and the ground state solution of the functional Schr\"{o}dinger
equation is written as
\begin{eqnarray}
\Psi_{\mathbf k}^{\rm R,I}\left[\tau,\zeta^{\rm R,I}_{\mathbf k}\right]=\sqrt{N_k(\tau)}\exp\left(-\frac{\Omega_k(\tau)}{2}\left(\zeta_{\mathbf k}^{\rm R, I}\right)^2\right).
\label{psi-ri}
\end{eqnarray}
Here $\Omega_k$ is related to the mode functions $f_k$ in the
Heisenberg picture as
\begin{eqnarray}
\Omega_k=-i\frac{f_k^{*'}}{f_k^*}.
\label{Omega1}
\end{eqnarray}
Thus, in Schr\"{o}dinger picture the power spectrum of $\zeta$ turns
out to be
\begin{eqnarray}
\mathcal{P}_{\zeta}(k)=\frac{k^3}{2\pi^2}|f_k|^2=\frac{k^3}{2\pi^2{\rm Re}\,\Omega_k},
\label{power-zeta}
\end{eqnarray}
and thus the power spectrum of the comoving curvature perturbations can
be obtained as
\begin{eqnarray}
\mathcal{P}_{\mathcal{R}}(k)=\frac{k^3}{8\pi^2\epsilon M^2_{\rm Pl}}\frac{1}{a^2{\rm Re}\,\Omega_k}.
\label{power-R}
\end{eqnarray}

It has been proposed in \cite{Martin:2012pea} that CSL-like
modifications can be applied to the inflationary perturbations directly
in the Fourier space by adding CSL-like non-linear and stochastic
terms to the functional Schr\"{o}dinger equation of $\zeta$ which then
looks like 
\begin{eqnarray}
d\Psi_{\mathbf k}^{\rm R,I}&=&\left[-i\hat{\mathcal{H}}_{\mathbf k}^{\rm R,I}d\tau+\sqrt{\gamma}\left(\hat\zeta_{\mathbf k}^{\rm R,I}-\left\langle\hat\zeta_{\mathbf k}^{\rm R,I}\right\rangle\right)dW_\tau\right.\nonumber\\
&&\left.-\frac{\gamma}{2}\left(\hat\zeta_{\mathbf k}^{\rm R,I}-\left\langle\hat\zeta_{\mathbf k}^{\rm R,I}\right\rangle\right)^2d\tau\right],
\label{func-sch-eq}
\end{eqnarray}
where the stochastic behavior due to CSL mechanism is encoded in the
Wiener process $W_\tau$ and $\gamma$ is called the collapse parameter. The most
general stochastic wave-functional which satisfies this stochastic
functional Schr\"{o}dinger equation can be written as
\begin{eqnarray}
&&\Psi^{\rm R,I}_{\mathbf k}\left(\tau,\zeta_{\mathbf k}^{\rm R,I} \right)=\left|\sqrt{N_k(\tau)}\right|\exp\left\{i\sigma_{\mathbf k}^{\rm R,I}(\tau)+i\chi_{\mathbf k}^{\rm R,I}(\tau)\zeta_{\mathbf k}^{\rm R,I}\right.\nonumber\\
&&\left.-\frac{{\rm Re}\,\Omega_k(\tau)}{2}\left[\zeta_{\mathbf k}^{\rm R,I}-\bar\zeta_{\mathbf k}^{\rm R,I}(\tau)\right]^2-i\frac{{\rm Im}\,\Omega_k(\tau)}{2}\left(\zeta_{\mathbf k}^{\rm R,I}\right)^2\right\},\nonumber\\
\end{eqnarray}
where $\bar\zeta_{\mathbf k}^{\rm R,I}$, $\sigma_{\mathbf k}^{\rm
  R,I}$ and $\chi_{\mathbf k}^{\rm R,I}$ are real numbers.

In \cite{Martin:2012pea}, the collapse parameter $\gamma$ was taken to
be constant and it was inferred that such a case is not capable of
explaining the quantum-to-classical transition of primordial
modes. Then it was argued in \cite{Das:2013qwa} that taking $\gamma$
to be constant the scenario loses one of the crucial features of CSL
dynamics, known as amplification mechanism. In CSL-modified quantum
mechanics, the collapse parameter is taken to be directly proportional
to the mass of the system and its number density reflecting the fact
that heavier objects become classical faster than the lighter
ones. Similarly, in tune with the expectation that superhorizon modes
behave classically one could assume in the cosmological context that
the collapse parameter would become stronger as a generic mode starts
to cross the horizon facilitating its collapse to one of its field
eigenstates. Hence a phenomenological form of the collapse parameter
was proposed in \cite{Das:2013qwa} as
\begin{eqnarray}
\gamma=\frac{\gamma_0(k)}{(-k\tau)^\alpha},
\label{time-gamma}
\end{eqnarray}
where $0<\alpha<2$. \footnote{Strictly speaking, taking $\gamma$ to
  depend on k is an assumption - it is by now means obvious that this
  follows from the original CSL equation. As such, we do not at
  present know what the form of the field-theoretic CSL equation in a
  curved space time should be. The arbitrariness in the choice of
  $\gamma$ gets constrained by the consideration of tensor
  perturbations, as we demonstrate below, and this is the key point of
  the paper. We make the reasonable assumption that $\gamma$ is the
  same for scalar and tensor perturbations.} It was also shown in
\cite{Das:2013qwa} that with $1<\alpha<2$ the CSL-modified scalar
dynamics successfully explains the quantum-to-classical transition of
primordial scalar modes without destroying the phase coherence of the
superhorizon modes essential to explain the peaks and troughs of the
CMBR anisotropy spectrum.

Now, let us calculate the power spectrum in this scenario. In such a
case it has been calculated in \cite{Das:2013qwa} that
\begin{eqnarray}
a^2{\rm Re}\,\Omega_k=\frac{k\gamma_0(k)}{H^2}(-k\tau)^{-(1+\alpha)},
\label{re-omega}
\end{eqnarray}
which indicates that the power spectrum would not be time-invariant on
superhorizon scale unlike the standard scenario. This feature has been
observed in both \cite{Martin:2012pea, Das:2013qwa} and as a cure to
it the power was calculated at the end of inflation and not at the
horizon-crossing of each mode by using
\begin{eqnarray}
-k\tau=\frac{k}{k_0}e^{-\Delta N},
\end{eqnarray}
where $k_0$ is the comoving wavenumber of the mode which is at the
horizon today $k_0=a_0H_0$ and $\Delta N$ is the number of efolds the
mode has spent outside the horizon after its exit and thus for
observationally relevant modes $\Delta N\sim50-60$. We would
henceforth consider $\Delta N\sim60$. Also to make the power spectrum
nearly scale-invariant we chose the scale-dependence of $\gamma_0(k)$
as
\begin{eqnarray}
\gamma_0(k)=\tilde\gamma_0\left(\frac{k}{k_0}\right)^\beta.
\end{eqnarray}
Then the power spectrum of comoving curvature perturbations turns out
to be
\begin{eqnarray}
{\mathcal P}_{\mathcal R}&=&\frac{1}{8\pi\epsilon M_{\rm Pl}^2}\frac{k_0^2H^2}{\tilde\gamma_0}e^{-(1+\alpha)\Delta N}\left(\frac{k_*}{k_0}\right)^{3+\alpha-\beta}\left(\frac{k}{k_*}\right)^{3+\alpha-\beta}\nonumber\\
&\equiv&A_s(k_*)\left(\frac{k}{k_*}\right)^{3+\alpha-\beta}.
\label{power-scalar-csl}
\end{eqnarray}
Incorporating the correction to the tilt in the spectrum due to
quasi-de Sitter evolution of the background, the power spectrum gets
modified to
\begin{eqnarray}
{\mathcal P}_{\mathcal R}=A_s(k_*)\left(\frac{k}{k_*}\right)^{3+\alpha-\beta+2\eta-4\epsilon}= A_s(k_*)\left(\frac{k}{k_*}\right)^{n_s-1},
\label{power-scalar-csl1}
\end{eqnarray}
where the power is now calculated at the end of inflation. 
\subsection{Tensor perturbations}

The tensor perturbations would now be straightforward to calculate
once the scalar analysis is done as each of the helicity components of
the Fourier tensor mode behaves like massless scalar perturbations, as
discussed before. It should be noted that tensor perturbations are
gauge-invariant by construction and the redefined tensor perturbations
$v_k^s$ of Eq.~(\ref{tensor-conf}) can be identified as the
Mukhanov-Sasaki variable of scalar perturbations defined in the
previous section.  Hence in the Schr\"{o}dinger picture each helicity
component $v_k^s$ of the tensor modes can be expressed as functionals
given in Eq.~(\ref{scalar-func}) following similar functional
Schr\"{o}dinger equation given in Eq.~(\ref{sch-eq}) in the standard
scenario. The ground state solutions of the functional Schr\"{o}dinger
equation would also be Gaussian as given in Eq.~(\ref{psi-ri}).
Similarly the power spectrum of $v_k^s$ would be same as of $\zeta$
given in Eq.~(\ref{power-zeta}) and can be written as
\begin{eqnarray}
\mathcal{P}_{v^s}(k)=\frac{k^3}{2\pi^2{\rm Re}\,\Omega_k},
\end{eqnarray}
and following Eq.~(\ref{tensor-conf}) one can write down the power
spectrum for the tensor modes as
\begin{eqnarray}
  \mathcal{P}_h=\sum_s\frac{4}{a^2M_{\rm Pl}^2}{\mathcal P}_{v^s}=\frac{2}{\pi^2M_{\rm Pl}^2}\frac{k^3}{a^2{\rm Re}\,\Omega_k}.
\end{eqnarray}

At this point, we assume CSL collapse mechanism affects each helicity
mode of the gravitons the same way as it affects the inflatons. This
is the simplest scenario to imagine and is in conformity with the
philosophy that the collapse mechanism should be universal in nature.
Since in our way of implementing the collapse mechanism we essentially
use the fact that each mode is an independent harmonic oscillator to
modify the equation of motion, there is no reason to expect that this
modification should be sensitive to details of the underlying nature
of the field. Hence the CSL-modified dynamics of each helicity mode of
the gravitons would be same as that of the massless inflatons or the
gauge-invariant Mukhanov-Sasaki variable which has been considered in
the previous section. For redefined tensor modes $v_s$, ${\rm
  Re}\,\Omega_k$ would have the same form as given in
Eq.~(\ref{re-omega}), following which the power spectrum of the tensor
modes can be determined as
\begin{eqnarray}
\mathcal{P}_h&=&\frac{2}{\pi^2M_{\rm Pl}^2}\frac{k_0^2H^2}{\tilde\gamma_0}e^{-(1+\alpha)\Delta N}\left(\frac{k_*}{k_0}\right)^{3+\alpha-\beta}\left(\frac{k}{k_*}\right)^{3+\alpha-\beta}\nonumber\\
&\equiv& A_T(k_*)\left(\frac{k}{k_*}\right)^{3+\alpha-\beta},
\label{csl-tensor}
\end{eqnarray}
and considering the tilt in the power due to quasi-de Sitter
background evolution one gets 
\begin{eqnarray}
\mathcal{P}_h=A_T(k_*)\left(\frac{k}{k_*}\right)^{3+\alpha-\beta-2\epsilon}=A_T(k_*)\left(\frac{k}{k_*}\right)^{n_T}.
\label{csl-tensor1}
\end{eqnarray}

\subsection{Observables}

Let us now illustrate how the CSL-modified primordial dynamics differ
from the standard one observationally. The first thing to note from
Eq.~(\ref{power-scalar-csl}) and Eq.~(\ref{csl-tensor}) is that the
tensor-to-scalar ratio remain the same in the modified dynamics:
\begin{eqnarray}
r=16\epsilon,
\end{eqnarray}
which indicates from the Lyth bound that the field excursion during
inflation would be super-Planckian even in this case. The scalar
spectral index and the tensor spectral index would now become (from
Eq.~(\ref{power-scalar-csl1}) and Eq.~(\ref{csl-tensor1}))
\begin{eqnarray}
n_s-1&=&\delta+2\eta-4\epsilon,\nonumber\\
n_T&=&\delta-2\epsilon,
\end{eqnarray}
where we have defined $\delta=3+\alpha-\beta$. We note here that,
although the model has three free parameters $\tilde\gamma_0$,
$\alpha$ and $\beta$ to begin with, the modification to spectral
indices can be captured in one effective free parameter $\delta$. The
observation of the scalar spectral index by PLANCK indicates that the
quantity $\delta$ can at best be of the order of slow-roll parameters
so that the comoving curvature power spectrum remains to be nearly
scale-invariant.

We also note from Eq.~(\ref{csl-tensor}) that the tensor amplitude in
such a case does not remain to be a sole function of Hubble parameter,
the prime feature which is used to determine the scale of inflation
using the BICEP2 observations. Even though, if we consider that
inflation has indeed taken place at that high scale, then that would
help estimating the collapse parameter as
\begin{eqnarray}
\tilde\gamma_0\sim k_0^2e^{-(1+\alpha)\Delta N},
\end{eqnarray}
which turns out to be extremely small. A stronger collapse mechanism
can then bring down the scale of inflation even though the field
excursions would remain to be super-Planckian.

Most interestingly what this modified dynamics does is to change the
consistency relation of the single-field models. In such a scenario
the consistency relation turns out to be
\begin{eqnarray}
r=-8n_T+8\delta.
\end{eqnarray}
Hence independent accurate measurements of $r$ and $n_T$ would give us
a direct handle on $\delta$ in this model.
\section{Discussion and Conclusion}

Belief in quantum nature of primordial perturbations makes it
essential to understand the apparent classicality of inhomogeneity it
gives rise to. The squeezing of the superhorizon modes and decoherence
partially explain this conundrum. But the single outcome problem
cannot be addressed without appealing to either many-worlds
interpretation or collapse mechanisms. Collapse mechanisms generically
modify the dynamics of the primordial perturbations and so are
expected to leave their imprints on cosmological observables. In
\cite{Martin:2012pea, Das:2013qwa} some toy models of CSL applied to
inflationary dynamics were explored. In \cite{Das:2013qwa}, within the
context of an illustrative example, it was shown that classicality of
perturbations can be achieved while still preserving scale-invariance
and phase coherence for a certain parameter range. In the light of
BICEP results which report high enough tensor-to-scalar ratio,
inconsistent with purely classical dynamics of the primordial tensor
modes \cite{Ashoorioon:2012kh, Bose:2001fa, Krauss:2013pha,
  Krauss:2014sua}, we revisit this model to investigate its
observational consequences.

The first point to note is that in this illustrative example the
tensor-to-scalar ratio remains unchanged indicating super-Planckian
field excursions during inflation as demanded by Lyth bound
\cite{Lyth:1996im}. Secondly, the tensor amplitude no longer directly
yields the scale of inflation as the collapse parameter also enters in
this conversion. A stronger collapse parameter would bring down the
scale of inflation. Unfortunately, so far there is no other
theoretical or observational guide to estimate the scale of collapse
parameter. Furthermore, the spectral tilts get modified by the one and
the same combination of the free parameters of the model which
eventually also change the consistency relation of the standard
single-field scenario by capturing the deviation from this in one
effective free parameter $\delta$. From PLANCK's observation of the
scalar spectral index, the free parameter $\delta$ can at best be of
the order of slow-roll parameters. It is also in principle possible
that $\delta$ be identically zero in which case the collapse mechanism
achieves the required classicality without leaving any imprint on
observations. All the same, it is more reasonable to expect $\delta$
to be non-zero which then would make this model testable by
observations.  One important feature which distinguishes this kind of
modification from other scenarios which also modify the consistency
relation like curvaton \cite{Fujita:2014iaa} or multifield
\cite{Kim:2006ys} models is that generically such extensions of the
minimal model modify the scalar sector while leaving the tensor sector
untouched, however the collapse models modify both and hence are
potentially distinguishable by precision measurement of $n_T$. However
modification to initial conditions of tensor modes, i.e. deviation
from Bunch-Davies vacuum would also modify the tensor spectral tilt
\cite{Ashoorioon:2014nta} reflecting the scale-dependence of
Bogoliubov coefficients. It is possible to arrange this
scale-dependence in such a way as to shift the tensor spectral index by
a constant, thus mimicking the effect of the collapse mechanism
considered here. But in principle, the generic scenario of non-Bunch
Davies initial condition should be distinguishable from the collapse
scenario. This would require precision measurement of running of
tensor spectral index which has been argued to be possible in the near
future \cite{Caligiuri:2014sla}.
\begin{acknowledgements}
The work of TPS is supported by a John Templeton Foundation grant
[39530]. SD and SS would like to thank Kinjalk Lochan for useful
discussions. 
\end{acknowledgements}

\end{document}